\documentclass[11pt]{article}
\usepackage[a4paper,hmarginratio=1:1,vmarginratio=2:3,totalwidth=16.5cm,
totalheight=21.6cm]{geometry}  
\usepackage{epsfig} 
\newcommand{\ra}{\rightarrow}
\newcommand{\cJ}{{\cal J}}
\newcommand{\cO}{{\cal O}}
\newcommand{\cE}{{\cal E}}
\newcommand{\cR}{{\cal R}}
\newcommand{\cG}{{\cal G}}
\newcommand{\bZ}{{\bf Z}}
\newcounter{oldcounter} 
\addtocounter{equation}{1}
\begin{document} 
\begin{flushright}  
{DAMTP-2003-127}\\
\end{flushright}  
\vskip 3 cm
\begin{center} 
{\Large 
{\bf Compact Dimensions and their Radiative Mixing.}}\\
\vspace{1.9cm} 
{\bf D.M. Ghilencea}\\
\vspace{1.cm} 
{\it D.A.M.T.P., C.M.S., University of Cambridge, \\
Wilberforce Road, Cambridge CB3 OWA, United Kingdom}\\
\vspace{2.6cm}
\end{center} 

\begin{abstract}
For one and two dimensional field theory orbifolds we compute in the DR 
scheme  the full dependence on  the  momentum scale ($q$) of the
one-loop  radiative  corrections to the 4D gauge coupling. Imposing
the  discrete shift symmetry  of summing the infinite towers of
associated  Kaluza-Klein (KK) modes, it is shown  that higher
dimension operators are radiatively  generated as one-loop
counterterms for the case of two  (but not for one) compact
dimension(s).  They emerge as a ``radiative mixing'' of 
effects (Kaluza-Klein infinite sums) associated with both compact dimensions.
Particular attention is paid to the link of the one-loop corrections  with  
their counterparts  computed  in infrared 
regularised 4D N=1  heterotic string orbifolds  with N=2 sectors. 
The correction from these sectors usually ignores  higher order terms 
in the IR string  regulator ($\lambda_s\ra 0$) of type $\lambda_s \ln
\alpha'$, ($\alpha'\not= 0$) but  these become relevant in  the field
theory limit.  Such terms ultimately re-emerge in pure field theory 
calculations of $\Pi(q^2)$ as higher dimension one-loop counterterms. 
We stress the importance of such terms for the unification of gauge
couplings and for the predicted value of the string scale.

\vspace{0.5cm}
\end{abstract}

\newpage
\section{Radiative corrections to gauge couplings.}
One-loop radiative corrections to the 4D gauge couplings induced by
compact dimensions were extensively studied in the past.  In general
in a 4D renormalisable model such as the Standard Model (SM) or the
Minimal Supersymmetric Standard Model (MSSM), the one-loop ``running''
of the gauge couplings is logarithmic. If these models are considered
as low energy limits of higher dimensional models, additional
corrections to this ``running'' exist. These are
associated with compact dimensions and induced by the corresponding
Kaluza-Klein (KK) states which are charged under the gauge group of
the model. Such corrections were analysed in effective field theory
(see for example \cite{Dienes:1998vg,Oliver:2003cy,Goldberger:2002hb,
Lanzagorta:1995gp}) and in string theory 
models~\cite{Kaplunovsky:1992vs,Dixon:1990pc2,Mayr:1993mq}.

\begin{figure}[t]
\centerline{\psfig{figure=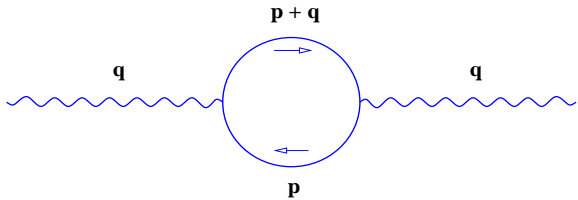,height=1.in,width=3.in,angle=0}}
\caption{\small One-loop diagram contributing to the gauge couplings,
with a fermion of mass $M_{\underline n}$
and its associated Kaluza-Klein tower in the loop. Its
expression $\Pi_{\mu\nu}(q^2) =\Pi(q^2)(q_\mu q_\nu-g_{\mu\nu}q^2)$
for $q^2\not=0$ can be read from eq.(\ref{grl}) for one or two 
compact dimensions.}
\label{figure1}
\end{figure}

In an effective field theory model  with one or two
additional compact dimensions one can compute the one-loop correction 
to the 4D gauge coupling by summing up individual contributions of 
the Kaluza-Klein states in the loop (Figure \ref{figure1}). 
 The correction is usually evaluated on-shell $(q^2=~0)$ and this 
is particularly true for the  string calculations, which in a
 more general setup also include the additional effect of the winding
modes (if present).  
The coupling corrected by this  one-loop
threshold correction depends  on the UV regulator/cutoff which
provides  an indication of the UV behaviour of the theory.  
Effective field theory calculations of the one-loop
correction  $\Pi(q^2\!=\!0$)
\cite{dg1,Ghilencea:2002ff,Ghilencea:2003kt} show remarkable
quantitative agreement with heterotic string results at ``large''
compactification radii. See however \cite{Ghilencea:2002ak} for 
a further discussion on the link between these approaches.

The 4D gauge coupling obtained as above (hereafter denoted
 $\alpha(0)$) is usually regarded as the coupling at some ``high''
 (compactification) scale  \cite{Dienes:1998vg}.  Below the
 compactification scale it
 is usually assumed that a 4D theory and corresponding logarithmic
 ``running'' (in $q^2$) apply.  This is indeed the case under the
 assumption that the  massive Kaluza-Klein states
 decouple at a momentum scale $q$ above or of the order of the 
 compactification scale(s). In general such decoupling is true for 
a finite number of
 states. However, in the case of evaluating the contribution of 
 many  {\it infinite}-level 
 towers of Kaluza-Klein states such situation may turn
 out to be slightly different\footnote{At the technical level and from
 a 4D point of view this is related to whether all the series which sum
 Kaluza-Klein radiative effects from compact dimensions 
 are (uniformly) convergent and can be integrated term by term.}. 
 To illustrate this we use an effective field model to analyse  the
 more general case of $\Pi(q^2\not=0)$ for the one-loop correction
 (Figure \ref{figure1}).
 This will  reveal a new effect,  present  when summing 
 over  infinite towers of KK modes. In such case it turns out
 that higher dimensional operators are radiatively generated  as {\it one-loop
 counterterms} for the case of two  (but not for one) 
 compact dimension(s). This is a result of a (one-loop) ``mixing'' of the two
 contributions associated each with one compact dimension. 
 Such counterterms are not present if the KK towers
 are truncated to any large  number of modes.  
 We discuss in detail the link  of such higher 
dimension operators in our field theory approach 
with one-loop  heterotic string calculations
and their (dis)agreement.
Special attention is paid below  to the regularisation of the divergent
series of integrals involved, performed in a gauge invariant way.

To begin with, let us consider the general structure of the one-loop
correction in two simple 4D toy-models which have one and two
additional compact dimensions, respectively. We assume each model has
a gauge group $G$ with 4D tree level gauge coupling $\alpha$, and that
they are compactified on a one- and two-dimensional orbifolds
respectively.  For our discussion the exact details of
compactification are somewhat unimportant and one can work in the
setup presented in \cite{Dienes:1998vg}. 4D N=1 supersymmetry is a
necessary ingredient to ensure only wavefunction-induced corrections
to the 4D gauge coupling.  To illustrate the main
point one can  use the QED action in 5D and 6D respectively, to
 perform a one-loop calculation of the vacuum polarisation
diagram in Fig.~\ref{figure1} with a fermion in the loop and its
associated tower of KK states. The result obtained  is more general 
and applies to the non-Abelian case too.
We use the dimensional regularisation
scheme (DR) for the UV divergences.  Following standard calculations
(see Appendix A in \cite{Dienes:1998vg}), after
performing the traces over the Dirac $\gamma$-matrices and with the
notation $\Pi_{\mu\nu}(q^2)=\Pi(q^2)(q_\mu q_\nu-g_{\mu\nu}q^2)$ one
has\footnote{in the 't Hooft gauge \cite{Dienes:1998vg}.}
\begin{eqnarray}\label{grl}
\Pi(q^2)=  \alpha\, (2\pi)^\epsilon \,
\frac{\beta}{4\pi} \sum'_{\underline n} \int_{0}^{1} d x\, 6 \, x\,
(1-x) \, \Gamma[\epsilon/2]\,
\bigg[\frac{\mu^2} {\pi (M_{\underline n}^2 + x\, (1-x) \,
q^2 )} \bigg]^{\frac{\epsilon}{2}}
\end{eqnarray} 
Here $\beta$ is the one-loop beta-function coefficient 
of a state in the loop associated with a KK tower,
 $\alpha$ is the gauge coupling; $\mu$ is the usual
finite, non-zero mass scale introduced by the  dimensional 
regularisation scheme.
Eq.(\ref{grl}) is just the familiar 4D result  \cite{CL} for a state of mass
$M_{\underline n}$ in the loop,  with an additional sum over the KK levels
$\underline n$. 
The ``primed'' sum over $\underline n$ runs over all integers $\underline
n=n\in\bZ$ with $n\not =0$ for one compact dimension and $\underline
n=(n_1,n_2)$ with $n_{1,2}$ integers and $(n_1,n_2)\not=(0,0)$ for two
compact dimensions. We thus exclude this ``zero-mode'' contribution
since we are only interested in the effect of the {\it massive} 
Kaluza-Klein modes on the gauge coupling and their decoupling
at $q^2$ smaller than  the compactification (scales)$^2$.  
We also assumed that a discrete ``shift'' symmetry of the Kaluza-Klein 
modes/levels $n\ra n+1$ holds true, and this imposes the summation over the
whole, infinite KK tower(s). One has from eq.(\ref{grl})
\begin{eqnarray}\label{lprt}
\Pi(q^2)& =&\alpha\, (2\pi)^\epsilon \, \frac{\beta}{4\pi} \, \int_{0}^{1}
d x\, 6\, x\, (1-x) \, \sum'_{\underline n} \int_{0}^{\infty}
\frac{dt}{t^{1-\epsilon/2}} \,\, e^{-\pi \, t\, (M_{\underline n}^2 +x
(1-x)\, q^2)/\mu^2}
\end{eqnarray}
which simplifies if $q^2=0$,
\begin{eqnarray}
\Pi(0)&=&\alpha\,(2\pi)^\epsilon\frac{\beta}{4\pi} \sum_{\underline
n}'\int_{0}^{\infty} \frac{dt}{t^{1-\epsilon/2}} \,\, e^{-\pi \,
t\,M_{\underline n}^2/\mu^2}
\end{eqnarray}
Eq.(\ref{lprt}) gives the general structure of $\Pi(q^2)$ in models
with compact dimensions. The UV region $t\!\ra\! 0$ is DR regularised.
If $M_{\underline n}=0$ for some level $\underline n$, the exponent in
(\ref{lprt}) vanishes at $x=0,1$ and then an IR regulator at $t\!\ra\!
\infty$ is also needed. This is introduced by an ``infrared'' mass shift
$\lambda^2\!\ra\!0$ of masses $M_{\underline n}^2$, ensuring the
integral over $t$ is exponentially suppressed at $t\!\ra\! \infty$ for
 any $x\in [0,1]$.

$\Pi(0)$ was evaluated in many effective field theory models using UV
cutoff regularisation, see for example
\cite{Dienes:1998vg,Lanzagorta:1995gp}, but such
regularisations are not gauge invariant.  For generic orbifolds with 
two compact dimensions with/without Wilson lines, $\Pi(0)$ was computed in
\cite{Ghilencea:2002ff,Ghilencea:2003kt,Ghilencea:2002ak} where the
quantitative agreement with its heterotic string counterpart
\cite{Dixon:1990pc2,Mayr:1993mq} was discussed in 
detail\footnote{See ref.\cite{dg1} for a general field theory 
computation  of $\Pi(0)$ in DR, proper-time and zeta-function 
regularisations.}.
For one compact dimension $\Pi(q^2)$ was  computed  in the DR scheme in 
\cite{Goldberger:2002hb}. At this point we discuss separately 
the cases of one and two compact dimensions for $\Pi(q^2)$ to 
reveal an important difference.

\section{One compact dimension.}\label{section1}

Our calculation of $\Pi(q^2)$ for one compact dimension is different  from
that in \cite{Goldberger:2002hb}, and is performed here in a manner 
suitable to a later comparison with the case of two compact dimensions. 
To evaluate  $\Pi(q^2)$ we need to know the 4D
Kaluza-Klein mass spectrum. This depends on compactification details, 
but for our purpose  we use its most general structure 
\begin{equation}\label{wdrs}
M_n=\frac{1}{R^2} (n+\rho)^2 +\lambda^2
\end{equation}
$R$ is the radius of compactification and $\rho$ depends on the
orbifold twist or on some additional effects such as Wilson lines
vev's.  $\lambda$ may be due to massive initial 5D matter fields.
This formula applies for example to models with compactification 
on $S^1/Z_2$, $S^1/(Z_2\times Z_2)$. In some models $\lambda$ may 
actually vanish and if $M_n$ also vanishes for some value 
of $n$ (if $\rho$ is an
integer), the whole exponent in eq.(\ref{lprt}) vanishes for $x=0, 1$.  
Mathematical consistency of eq.(\ref{lprt}) then requires a
 mass shift of the {\it whole} tower (zero-mode included) 
by an infrared mass regulator,
so we would  need introduce  $\lambda\not=0$ and then take  $\lambda\ra 0$.
 For appropriate re-definitions of the parameters 
$\rho$, $\lambda$ and $R$, most cases of models  with  one extra 
dimension can be recovered. Here we keep $R,\rho,\lambda$ as arbitrary
parameters.

We use  eq.(\ref{wdrs}) in  eq.(\ref{lprt}) and 
the following result\footnote{
Adding a zero-mode contribution to eq.(\ref{drlast}) would cancel
the pole $1/\epsilon$ and the $\ln[\pi e^\gamma \tau
(\rho^2+\delta/\tau)]$ term.} in DR (see Appendix A of \cite{dg1})
\begin{eqnarray}\label{drlast}
 \int_{0}^{\infty} 
\frac{dt}{t^{1+\epsilon}} \sum'_{m\in\bZ}
            e^{-\pi\,t\, [\tau \,(m+\rho)^2 + \delta]}
= \frac{1}{\epsilon}-\ln\frac{\vert 2\, \sin
\pi(\rho+i(\delta/\tau)^{\frac{1}{2}})\vert^2}
{\pi e^\gamma\,\tau\, (\rho^2+\delta/\tau)}
,\qquad \delta\geq 0,\,\,\, \tau>0.
\end{eqnarray} 
With the notation $h(x)=x(1-x)$,\, $\sigma^2\equiv q^2\, R^2$ and
$\nu\equiv \lambda R$ we find 
from eq.(\ref{lprt}) to order $\cO(\epsilon)$
\begin{eqnarray}\label{dffrqrd}
\Pi(q^2)&=&\alpha \frac{\beta}{4\pi}\, 
\bigg\{
-\frac{2}{\epsilon}-\ln [4\pi e^{-\gamma}] +
6\int_0^\infty \! dx\,\,h(x)\nonumber\\
\nonumber\\
& \times&
\!\!\!\!\bigg[
\ln\frac{\rho^2+\nu^2+h(x)\, \sigma^2}{(R\mu)^2}
-
2\pi [\nu^2+h(x)\, \sigma^2]^{\frac{1}{2}}
-
\ln\bigg\vert 
1-e^{2 i\pi\rho}\,e^{-2  \pi (\nu^2+h(x)\, \sigma^2)^\frac{1}{2}}
\bigg\vert^2
\bigg]
\bigg\}
\end{eqnarray}
The dependence of the  couplings on $q^2$ is then
\begin{eqnarray}\label{prtgfbm}
\alpha^{-1}(q^2)-\alpha^{-1}(0)=\bigg[\Pi(q^2)-\Pi(0)\bigg] 
\, \alpha^{-1}(0)
\end{eqnarray}
The first two integrals in (\ref{dffrqrd}) give  logarithmic and 
linear terms in $q R$, depending 
on the relative size of the parameters
involved. The first integral may be regarded as
the contribution from a single state of mass equal to that of the 
zero-mode ($M_0$).

For our later  comparison 
with the two compact dimensions case it is important to notice
that the divergence $1/\epsilon$ cancels out in the difference 
$\Pi(q^2)-\Pi(0)$, to leave a  dependence  of the one-loop correction
on the parameters $q$, $R$ and $\lambda$ only. There are no 
terms in $\Pi(q^2)$ proportional to  $q^2/\epsilon$, which means
that higher dimensional (derivative) operators are not generated as
one-loop counterterms\footnote{They can however be generated beyond 
one-loop  level.}. The result for the change of the couplings 
with $q^2$ is then 
\begin{eqnarray}\label{wrdfghmnb}
\alpha^{-1}(q^2)-\alpha^{-1}(0) & =&  
\frac{\beta}{4\pi}\, (\cJ_1+\cJ_2+\cJ_3)
\nonumber\\
\nonumber\\
\cJ_1 &\equiv&
 \frac{4}{w} -\frac{5}{3}+
2 (w-2)(w+4)^{\frac12} w^{-\frac{3}{2}}
 \ln [((4+w)^{\frac12}-w^{\frac12})/2], 
\nonumber\\
\nonumber\\
\cJ_2 &\equiv& 
-\frac{3 \pi\sigma}{2}\bigg\{
 \bigg(\frac{\nu}{\sigma}\bigg)^3 
-\frac{7}{12} \bigg(\frac{\nu}{\sigma}\bigg)
+\frac{1}{8} \bigg[3 + 8 \bigg(\frac{\nu}{\sigma}\bigg)^2
 -16\bigg(\frac{\nu}{\sigma}\bigg)^4\bigg]
 \arctan\frac{\sigma}{2\nu}\bigg\}
 \nonumber\\
\nonumber\\
\cJ_3& \equiv &-6 \int_{0}^{1} dx\, h(x) 
\ln \bigg\vert
\frac{1-e^{2 i \pi \rho -2\pi 
 (\nu^2+h(x) \sigma^2)^{\frac{1}{2}}}}{
1- e^{2 i \pi \rho -2 \pi \nu}}\bigg \vert^2
\end{eqnarray}
where we used the notation 
$w\equiv q^2/M_0^2= {\sigma^2}/{(\rho^2+\nu^2)}$.
For $w\! \ll\! 1$, one has  $\cJ_1=w/5 + \cO(w^2)$; 
for $w\gg 1$, $\cJ_1= -5/3 + \ln w+\cO(1/w)$. 
Also for  $\sigma\ll 1$, and $\nu:$ fixed:
 $\cJ_2 = - (\pi/5) \,\sigma^2/\nu + \cO(\sigma^4)$.
If $\sigma$ is fixed and
 $\nu\ll 1$: 
$\cJ_2= - 9 \sigma\pi^2 /32  +2\pi\nu+\cO(\nu^2)$ with  
the first term giving a  ``power-like'' (linear) correction in 
the momentum scale   ($\sigma^2\sim q^2$)  which is important if 
 $q^2\!\geq\!1/R^2$.  $\cJ_3$ gives only a mild 
dependence on the momentum $q$  suppressed for $q^2\!\geq\! 1/R^2$. 
One may set $\lambda=0$ if the spectrum (\ref{wdrs}) of the model
considered requires it and if $\rho$ is
non-integer/non-zero. In such case only the term power-like 
in momentum survives  in $\cJ_2$. 
Eqs.(\ref{wrdfghmnb})  give the dependence of the couplings on the
scale $q^2$, which is different from that on the UV {\it cut-off scale}
considered in \cite{Dienes:1998vg}.
  The distinctive behaviour in $q^2$ as compared  
to the 4D  case may be used for phenomenology,
searches for  effects of an extra 
dimension or unification of gauge couplings in 
 models with a compact dimension. 
The only parameter in this correction is the  scale $1/R$;
there is no dependence on the UV regulator/cutoff
at one-loop~level.

\section{Two compact dimensions.}

The previous analysis can be repeated for two compact dimensions.
For the 4D toy-model with two additional compact 
dimensions the Kaluza-Klein mass  spectrum has the  general form
\begin{equation}\label{zzzzzz}
M_{m_1,m_2}^2=\frac{\vert m_2- U m_1 \vert^2}{(R_2 \sin\theta)^2}
\end{equation}
where we introduced the notation  $U\equiv U_1+i U_2$ with 
$U=R_2/R_1 \exp(i\theta)$. $R_i$ are the radii of the two compact 
dimensions. This mass formula can be generalised to $T^2/Z_N$ 
orbifolds without changing the conclusions below.

An important remark is in place here. The {\it total} correction
$\Pi(q^2)$  includes the contribution of the zero-mode $(0,0)$, 
in addition to 
that of non-zero modes given by eq.(\ref{lprt}). According to
(\ref{zzzzzz}) $M_{0,0}=0$ and for $x$ reaching its limits of
integration $x=0,1$ the contribution of the 
zero-mode\footnote{This is of
  the form given in eq.(\ref{lprt}) without the sum 
over the KK levels.}  to $\Pi(q^2)$  would have vanishing  
exponent under the integral over $t$. 
This integral    would then be divergent in the infrared ($t\!\ra
\!\infty$). A mass shift $M_{n_1,n_2}^2\ra M_{n_1,n_2}^2+\lambda^2$ is
necessary so that the
{\it total} expression $\Pi(q^2)$ including {\it massless} modes 
is well-defined  {\it before} splitting the contributions to
 $\Pi(q^2)$  into those due to massless and  massive modes, 
respectively.  In (\ref{lprt}) one sums over massive modes
 only and the integral over $t$ is indeed well defined for 
$t\ra \infty$ because $M_{m_1,m_2}\not=0$ if $(m_1,m_2)\not=(0,0)$. 
However, the  above discussion requires us to keep the IR
 regulator in the massive sector as well. In the following 
the exponential in (\ref{lprt}) will therefore be changed   
to include the (dimensionless) 
IR regulator $\lambda_0$ required by the massless modes 
\begin{eqnarray}\label{trgr}
e^{-\pi \, t\, ( M_{m_1,m_2}^2+x(1-x) q^2)/\mu^2}\ra 
e^{-\pi \, t\, [( M_{m_1,m_2}^2+x(1-x) q^2)/\mu^2+\lambda_0^2]},\qquad 
\lambda_0\!\ra\! 0, \,\, \lambda\equiv \mu  \lambda_0
\end{eqnarray}
with $\lambda$ the infrared mass scale associated with the
regulator $\lambda_0$.
This observation   is important because the  UV and IR 
regularisation
limits, $\epsilon\ra 0$ and $\lambda_0\ra 0$ respectively
may not ``commute'' in eq.(\ref{lprt}), even though 
this equation  only sums non-zero modes which have IR-finite 
contribution.

To evaluate  eq.(\ref{lprt}) we  use the following result in DR
\vspace{0.1cm}
\begin{equation}\label{wr1}
\!\!\!\!\int_{0}^{\infty}\!\! \!\frac{dt}{t^{1+\epsilon}}\sum'_{m_1,m_2} 
\,\, e^{-\pi\, t \, [ \tau \,\vert U m_1 - m_2\vert^2 + \, \delta]}
=\frac{1}{\epsilon}
+\frac{\pi \delta}{\epsilon}\frac{1}{\tau U_2}
-\ln\bigg[4\pi e^{-\gamma}\frac{1}{\tau}\,\vert \eta(U)\vert^4 \bigg]
+E\bigg(\frac{\delta}{\tau}\bigg); \quad  \delta\geq 0,\, \tau\!>0
\end{equation} 
with $U= U_1+i U_2$. Eq.(\ref{wr1})  is valid for $0\leq \delta \vert
U\vert^2/(U_2^2 \tau)\leq 1$,
$0\leq  \delta/(\tau U_2^2)\leq 1$ which are  sufficient conditions
only. The ``primed'' sum runs over all integers
$(m_1,m_2)$ except the level $(0,0)$ and  $\eta(U)$,
$E(\delta/\tau)$ are functions defined in the Appendix.
The function $E(y)$ is vanishing in the limit $y \ra 0$. 
The result has  divergences in $\epsilon$ from  $t\!\ra\! 0$ but
 there are no divergences in  $\delta$ when $\delta\ra 0$
because the integrand is always exponentially suppressed at
$t\!\ra\!\infty$ for $(m_1,m_2)\not=(0,0)$.
Note the emergence of the term proportional to
$\delta/(\tau \epsilon)$ in addition to\footnote{Adding a zero mode
  $(0,0)$ to eq.(\ref{wr1}) would cancel $1/\epsilon$, but would not cancel 
the term proportional to $\delta/\epsilon$.}
 $1/\epsilon$ and which will play an important role in the following. 
This is to be compared  to the  integral in eq.(\ref{drlast}) where no
such term is present. The difference is due to the presence of two sums
under the integral in eq.(\ref{wr1}) rather than only one as in the
one compact dimension case,  eq.(\ref{drlast}).

To compute $\Pi(q^2)$ we apply  the substitution (\ref{trgr}) 
in (\ref{lprt}) and then use eq.(\ref{wr1}). With the notation 
$\cR^2\equiv   R_1 R_2 \sin\theta$ and  retaining 
 terms to $\cO(\epsilon)$ one finds from (\ref{lprt})
\vspace{0.2cm}
\begin{equation}\label{ttt6}
\Pi(q^2)  =  \alpha
\frac{-\beta}{4 \pi} 
\bigg\{\frac{2}{\epsilon}
+ 2\pi \frac{(\lambda \cR)^2}{\epsilon} 
+ \frac{2 \pi}{5} \bigg[\frac{(q \cR)^2}{\epsilon}  
+  (q \cR)^2\ln2\pi\bigg]
+\ln \bigg[4 \pi e^{-\gamma}  \vert\eta(U)\vert^4\, U_2 \,(\mu
  \cR)^2\bigg]+\cG(q)
 \bigg\}
\end{equation} 
with the constraint
\begin{eqnarray}\label{cgrhtjk}
\lambda^2+ \frac{1}{4}\, q^2 \leq
\min \left\{\frac{1}{R_1^2},\frac{1}{R_2^2}\right\}
\end{eqnarray}
This   (sufficient) condition
is derived  from the validity of eq.(\ref{wr1}). In the limit of
``removing'' the infrared regulator
 one takes $\lambda\ra 0$ or $\lambda^2 \ll 1/R_{1,2}^2$ which leaves a
condition for the upper value of the momentum scale at which the
above result still applies.
In (\ref{ttt6})\, the function  $\cG(q)$  (analytic) also  depends 
on  $R_1$, $R_2$, $\lambda$,  but does not depend on the 
UV regulator  $\epsilon$. Its exact expression is not relevant in the
following and is given in the 
Appendix, eq.(\ref{gq1}). In $\cG$ we can safely 
remove\footnote{This means that the limit
$\lambda\!\ra\! 0$ in $\cG$ does not interfere with the 
$\epsilon$ dependence, already isolated in (\ref{ttt6}).} 
the  dependence on the IR regulator $\lambda$ 
($\lambda\ra 0$) to find the result of eq.(\ref{gq2}).

Note the  presence in $\Pi(q^2)$ of the term $(q \cR)^2/\epsilon$ 
which does not have a counterpart in the case of 
one compact dimension.
Obviously such term is missed when evaluating only $\Pi(0)$.
A somewhat similar term in $\Pi(q^2)$ is  $(\lambda
\cR)^2\!/\epsilon$, since  $\lambda^2$ and $q^2$ are on equal
footing in $\Pi(q^2)$ in  the exponent  under the integral 
over $t$, see eq.(\ref{lprt}) with the replacement 
(\ref{trgr}). Again, if one had $\lambda\!=\!0$ in the  
(IR finite) massive modes sector, this term would have been 
missed too. 

\noindent
Following the one compact dimension example, one could in principle 
write from eqs.(\ref{prtgfbm}), (\ref{ttt6}) 
\begin{equation}\label{wrtrtwndmb}
\alpha^{-1}(q^2)-\alpha^{-1}(0) 
 =   \frac{-\beta}{4\pi}\frac{2\pi}{5} \bigg[
\frac{(q \cR)^2}{\epsilon} 
+  (q\cR)^2\ln(2\pi)+\frac{5}{2\pi} \bigg(\cG(q)-\cG(0)\bigg)\bigg]
\end{equation}
Eq.(\ref{wrtrtwndmb}) shows that  the pole $1/\epsilon$
present in both $\Pi(q^2)$ and $\Pi(0)$ cancels out 
in their difference, similar to the case of one compact 
dimension. The same applies to the $q$-independent terms, 
in particular  to the term $(\lambda \cR)^2/\epsilon$ involving 
the IR scale $\lambda$. One is thus left with the $q^2$ 
dependent terms, and of these the most important is that 
proportional to $(q\cR)^2/\epsilon$. This term has no equivalent  
in the case of one compact dimension, see eq.(\ref{dffrqrd}), 
(\ref{wrdfghmnb}). For $q^2$ close to the compactification 
(scales)$^2$,  $1/R_1^2$ or $1/R_2^2$ the coupling has a pole.  
Even if $q^2\ll 1/R_1^2$ and 
$q^2\ll 1/R_2^2$, since $\epsilon\ra 0$, one cannot set this term 
to 0, and a ``non-decoupling'' effect of the KK modes is manifest.
Therefore the limit of  scales $q$
well below the compactification scales (hereafter referred to as 
"infrared") and the  UV regularisation limit $\epsilon\ra 0$
 do not commute.
As a result a  UV-IR ``mixing'' effect
(IR-finite, UV-divergent) exists due to the first
term in\footnote{The term $(\lambda \cR)^2/\epsilon$ present in
  $\Pi(q^2)$ or $\Pi(0)$ but not in their difference is itself a 
similar UV-IR contribution \cite{Ghilencea:2002ak}.} 
eq.(\ref{wrtrtwndmb}). The KK level $(0,0)$  - if included - 
cannot change this 
picture, because its contribution does not bring in a  
$\delta/(\tau \epsilon)$ term to eq.(\ref{wr1}) responsible for
$(q\cR)^2/\epsilon$ in eq.(\ref{ttt6}).

One concludes that in this regularisation set-up the 
Kaluza-Klein non-zero modes give an effect even at momentum scales 
well below the compactification scale, where one would
 expect them to be decoupled. 
The presence of the UV-IR mixing term is a result of 
considering the effect of an infinite (rather than a ``truncated'') 
tower of Kaluza-Klein modes, and as a consequence such ``non-decoupling'' 
effect, induced by infinitely many 
modes, may not be unexpected in the
end. It  is then puzzling why the term $(q \cR)^2/\epsilon$
has no counterpart in the one compact dimension case,
 where we also summed over the whole KK tower. 
How can we explain this difference? As we discuss later, such term
corresponds to a counterterm  in the action 
$\cR^2 D_M F^{MN} D^K\! F_{KN}$  which 
cannot be generated in 5D  at one-loop \cite{Oliver:2003cy}
due to Lorentz invariance.
At the technical level one can   show that $q^2 \cR^2/\epsilon$ 
emerges as a  one-loop ``mixing'' of the effects
of two compact dimensions: it  arises as a mixed contribution between  a 
sum over "original" Kaluza-Klein  modes associated with one compact 
dimension and a  "Poisson re-summed" (or winding) 
zero-mode\footnote{Poisson re-summation in one dimension gives:
$\sum_{n\in\bZ} \exp (-\pi t\, n^2/R^2)=R/\sqrt t 
\sum_{p\in\bZ}\exp(-\pi p^2 R^2/t);$ here $n$ 
labels original KK modes while $p$ denotes their ``Poisson re-summed'' or
dual (winding) modes referred to in the text.}
 of a sum corresponding to  the  second compact dimension.  
It is then clear  why such term cannot  appear in the case of a single
compact dimension. 
 This shows explicitly a different behaviour of 
the radiative corrections with respect to the character even/odd of 
the number of  compact dimensions \cite{Candelas:ae} and  brings  
additional effects to those  discussed 
in previous works~\cite{Dienes:1998vg,Lanzagorta:1995gp}.

An immediate question is the regularisation dependence of the
existence of the term  $(q \cR)^2/\epsilon$.
Our comparative analysis shows that the effect exists for 
two compact dimensions but there is no counterpart for one compact 
dimension  where the same UV regularisation was used. This
gives some indication that the existence of the
 term $(q \cR)^2/\epsilon$  is not
the result of a particular UV regularisation choice.
 Further, our previous discussion on the IR regularisation does 
not affect the existence of this term, and finally, the DR 
scheme used is supposed to provide a UV well-defined and manifestly  
 gauge invariant  framework~\cite{Cheng:2002iz}. One may argue that the 
UV regularisation must not affect
the IR regime of the theory and that the DR scheme used in this
calculation might not respect this condition.
However, calculations closely related \cite{Ghilencea:2002ak}
using an UV regularisation with a proper-time cutoff 
($t\geq 1/\Lambda^2$) in eqs.(\ref{lprt}), (\ref{wr1}) instead of DR, 
yield a similar UV-IR ``mixing'' term\footnote{Eq.(\ref{wr1}) 
with UV cutoff regularisation instead of DR
has  $\pi \delta/(\epsilon \tau U_2)$ replaced by a term proportional to 
$\delta\ln\Lambda$ \cite{Ghilencea:2002ak}.}
  $(q \cR)^2 \ln \Lambda$, with the 
$1/\epsilon$ factor simply replaced by the logarithm of 
the UV cutoff $\Lambda$.

Eqs.(\ref{ttt6}), (\ref{wrtrtwndmb}) simply tell us that higher dimension
(derivative) operators need to be included for a fully consistent 
one-loop calculation. This is a significant difference from the 
 previous case of one compact dimension only. Indeed,
the presence of the term $q^2/\epsilon$ in the effective field theory
result shows  that for two compact dimensions  the DR regularisation 
with minimal subtraction is not sufficient and that higher 
dimensional operators are radiatively generated/required 
 as {\it one-loop counterterms}.  One such counterterm is 
$\cR^2 D_M F^{MN} D^K\! F_{KN}$ (for related discussions on this issue
 see Section IV B in~\cite{Oliver:2003cy}).
This is important for it establishes a direct link between the effects of 
two compact dimensions or their associated  infinite KK sums,  
and the role of  higher dimensional   operators.
In the absence of additional constraints to fix the (otherwise
arbitrary)  coefficient of
such counterterms, the corrections they induce will depend on it
with  implications for the predictive power of the models. 
In the case of  KK towers ``truncated''  to a large but
finite number of  KK modes, such counterterms are not
radiatively generated\footnote{For more details on the 
decoupling of infinitely many  modes in a $\lambda \phi^4$ theory see
\cite{Kubyshin:su}.}.

We  do not  address in the following the detailed  
implications for field theory of such higher dimensional 
operators, but discuss instead the  origin  of $(q\cR)^2/\epsilon$ 
or equivalently  $(q \cR)^2 \ln \Lambda$ in $\Pi(q^2)$, from a  
{\it heterotic} string
perspective. This is important because it will show the link between 
the   higher dimensional operators as  one-loop counterterms 
in the field theory approach to $\Pi(q^2)$ 
 and the  one-loop radiative effects in string\footnote{This can be
done even  though the string  only  computes $\Pi(0)$ rather than 
$\Pi(q^2)$,  see later.}.
In doing so we consider that the string provides a ``UV completion''
of the  field theory case, 
with the latter  recovered in the limit $\alpha'\ra 0$ of the string, 
as shown  in \cite{Ghilencea:2002ff,Ghilencea:2003kt,Ghilencea:2002ak}
(also \cite{dg1}).
A string counterpart of the one-loop correction to gauge couplings 
considered above is that induced  by the N=2 sectors  of 4D N=1 toroidal 
orbifolds.   Such two-dimensional 
sectors bring one-loop corrections to the gauge couplings due 
to massive Kaluza-Klein  and winding states 
\cite{Kaplunovsky:1992vs,Dixon:1990pc2,Mayr:1993mq}. 
The (field theory limit of such)  string calculation for $\Pi(0)$  does 
agree with the pure field theory result for  $\Pi(0)$
\cite{Ghilencea:2002ff} which sums  Kaluza-Klein effects only, 
although  the relation between these different approaches is rather 
subtle \cite{Ghilencea:2002ak}. This is  particularly true 
when analysing the more  general case of $\Pi(q^2)$. Let us explain  this 
in detail.

The one-loop string calculation for $\Pi(0)$
\cite{Kaplunovsky:1992vs,Dixon:1990pc2}
which sums only massive modes' effects
needs itself a regularisation, this time in the IR region only.
In  string theory  one ultimately computes a one-loop diagram 
associated with $\Pi(0)$ rather than $\Pi(q^2)$ which we would 
need for comparison with  eq.(\ref{ttt6}). 
However, since $q^2$ and $\lambda^2$ are on equal
footing\footnote{By this we mean that in equation (\ref{ttt6}) 
there are both  $(\lambda\cR)^2/\epsilon$ and $(q \cR)^2/\epsilon$ terms.}
 in  $\Pi(q^2)$ of eq.(\ref{ttt6}) and also in 
the exponential in  (\ref{lprt}) with replacement
 (\ref{trgr}), it is enough to investigate the role of the string 
counterpart of our $\lambda$. This is just  the IR regulator in
string  (hereafter  denoted $\lambda_s$) which, unlike $q^2$, is 
also present in $\Pi(0)$  computed by string, and can still convey 
some information  about $\Pi(q^2\not=0)$~!

The IR regularised  string result for $\Pi(0)$ contains in addition to 
the well-known one-loop result~\cite{Dixon:1990pc2}, higher 
order terms in the IR regulator which in a DR
scheme  of the IR divergence have for example, the form\footnote{In a 
modular invariant IR  regularisation of the string
such $\alpha'$-dependent terms should be $SL(2,Z)_T$ invariant.}
$\lambda_s\ln\alpha'$. For technical details on how such term can arise
in string, from the degenerate orbits of the modular group $SL(2,Z)$,
see for   example Appendix\footnote{See eq.(A-1),(A-10),(A-12)
in \cite{ks}. (A-12) brings $\cO(\varepsilon)$  term 
$\varepsilon\ln (T_2 U_2)$, ($T_2\!\!\sim\!\!{R_1 R_2}/{\alpha'}$) 
discussed here with $\varepsilon\!\ra\! \!\lambda_s$.} A of
ref.\cite{ks} and also  the calculation in the Appendix of 
\cite{Dixon:1990pc2}.  
Here the IR  string regulator  $\lambda_s\!\ra\! 0$  and 
$\alpha'\sim 1/M_s^2$ with $M_s$ the string scale. For 
$\alpha'\!\not=\!0$ the term $\lambda_s\ln\alpha'$ 
 vanishes when $\lambda_s\ra 0$ and this explains why  it is not 
kept in the final,  infrared  regularised string result.

What does this tell us for the pure field
theory approach to $\Pi(0)$ or $\Pi(q^2)$ 
which sums KK effects only?  In the
field theory limit of the string calculation, one takes 
$\alpha'\!\ra\! 0$ (infinite string scale) to
suppress  string effects (winding modes) but keep those due to {\it
  massive} KK states only, considered in field theory. 
In such  case, the value of $\lambda_s\ln\alpha'$ 
depends on the order of taking the limits of IR regularisation 
$\lambda_s\ra 0$ and of field theory $\alpha'\!\ra\! 0$. 
This situation applies to other IR 
regularisations \cite{Dixon:1990pc2,Mayr:1993mq} of the string as well.
We are not aware of any string symmetry which imposes the order to take
these limits. The  term  $\lambda_s\ln\alpha'$   then becomes relevant 
in the  field theory limit. In this limit,  $\lambda_s$ 
($\lambda_s\!\ra\! 0$) is replaced by its field theory counterpart 
$\lambda^2$ ($\lambda^2\!\ra\! 0$)  
while $\alpha'$ plays the role that the  UV proper-time cutoff 
regulator $1/\Lambda^2$ does in the field theory approach. 
With these replacements, an UV-IR ``mixing'' term (IR finite, UV divergent) 
should emerge,  $(\lambda \cR)^2\ln \Lambda$, 
just as we found in  the field theory approach for $\Pi(0)$.
But this also tells us something about $\Pi(q^2)$ in field theory.
With the observation that $\lambda$  and $q$ are  on equal footing 
in $\Pi(q^2)$, this ``mixing'' terms implies 
that  one should  expect  in the field theory limit a term
$(q \cR)^2\ln \Lambda$ in the proper-time regularisation of the UV
or $ (q \cR)^2/\epsilon$ in the  DR scheme. 
This is in agreement with  our field theory result 
eq.(\ref{ttt6}) where such a term is found, and a strong 
consistency check of the field theory  calculation. 

This discussion  provides an insight into the role that 
higher dimension operators play in understanding the link between 
the {\it infrared} regularised string result
and pure field  theory  approaches for $\Pi(q^2)$. 
It implies in addition that corrections to gauge
couplings from    infrared regularised 
string calculations should retain the terms of structure
$\lambda_s\ln\alpha'$  in the final correction to $\Pi(0)$, if an 
{\it exact} agreement with their field theory counterpart is to be maintained.

This discussion  has  implications for the 
unification of gauge couplings in 4D supersymmetric models. 
We refer here to the attempts to  match the MSSM unification scale
with the  (heterotic) string scale value.  In MSSM-like models 
gauge couplings unify at  $\sim 2 \times 10^{16}$ GeV
\cite{Ghilencea:2001qq} which is
marginally below the predicted string scale 
$\sim g_{GUT}\, 5.27 \times 10^{17}$ 
GeV \cite{Kaplunovsky:1992vs}.  Our discussion on the
heterotic string  shows that for the models addressed 
the  effects  of higher dimension counterterms are  not included 
in  the one-loop string corrections. As a result
the predicted value  of the string scale $M_s$  does {\it not} 
include the effects from such operators.  
This finding should be considered  when attempting  solutions
for an {\it exact} matching
of the MSSM unification scale with the heterotic string scale.

\section{Final remarks and Conclusions}
For one and two dimensional orbifold compactification  we 
considered the  general case of evaluating at one loop level
$\Pi(q^2)$ in a manifestly gauge invariant scheme (DR).  
For these models we discussed comparatively the dependence  
of the couplings on the momentum scale $q^2$ and $1/R^2$, 
and the role of higher dimensional operators as one-loop counterterms.
These can be generated  when the summation over the {\it infinite} 
towers of Kaluza-Klein modes is performed. The analysis showed a 
different behaviour  of the one-loop correction with respect 
to the character even/odd  of the number of compact dimensions, 
with such operators generated for the case of  two but not for 
one compact dimension(s).

For one compact dimension  the change of 
the couplings $\alpha^{-1}(q^2)-\!\alpha^{-1}(q^{' 2})$ 
with respect to the momentum scale  is UV regulator independent  
at  {\it one-loop} level, unlike  the case of more common
approaches using cutoff regularisation \cite{Dienes:1998vg}. 
For one compact dimension the  results  can be used  
for phenomenology, unification of the gauge couplings and 
searches for effects from compact dimensions.

For two compact dimensions a similar analysis of the one-loop effects
suggests the existence of a correction which couples low
(``infrared'') scales below the compactification
scales,  to UV divergent terms.  This implies the existence
in this toy-model of some ``non-decoupling''  effects at low
energies,   due to a  ``mixing'' of the two {\it infinite} towers 
of  Kaluza-Klein states.
The emergence of such non-decoupling term  in the effective field
theory can be re-interpreted and explained simply by the 
presence - for two compact dimensions - of higher dimensional 
operators which are required as {\it one-loop  counterterms}.

We investigated in detail  the origin of such operators
from the  heterotic string perspective.  The origin of these
counterterms can  be related to string  corrections to $\Pi(0)$
of type $\lambda_s \ln \alpha'$ (with $\lambda_s\ra 0$ the IR string
regulator) which are  usually discarded  in the final
  one-loop string result, since $\alpha'\not= 0$.  
However, they become  relevant in the field 
theory limit, and also in pure field theory calculations
 where the two  regularisation limits (in IR, UV)  do not commute. 
This raises some intriguing issues  about
the {\it infrared} problem in heterotic string  and  its
link with higher dimensional  one-loop  counterterms in field theory.

If the Kaluza-Klein towers are ``truncated'' to a finite 
number of modes, such operators are not generated. In such 
case the discrete ``shift''  symmetry of summing over an infinite tower 
of Kaluza-Klein modes is broken. 
Under our initial assumption that such symmetry holds, 
the higher  dimensional operators  can  be seen to account for 
non-perturbative  effects. This is because  such operators are
ultimately related to effects of a zero-''mode'' of a  ``Poisson
re-summed'' Kaluza-Klein series, i.e. a winding mode 
(non-perturbative) effect. 

It is possible  that in fully specified models symmetry arguments 
may be identified to avoid the  presence of such higher dimension
operators.  Nevertheless we  think  these findings are important 
for phenomenology,  in particular for the scale of  unification
of gauge couplings. We argued that one-loop effects 
from higher dimension  counterterms are  not included in the
(predicted) value of the string scale and this has implications for its 
mismatch with the MSSM unification scale.

\vspace{0.7cm}
\noindent
{\bf Acknowledgements:}
The author thanks Graham Ross and 
Fernando Quevedo for  useful  discussions  on this problem. 
He also thanks the Theory Group at the University of Bonn  where 
this work was completed, for their hospitality and for many interesting
discussions on related topics. This work was supported by a
post-doctoral  research  fellowship from PPARC (UK).

\section*{Appendix}
\setcounter{equation}{0}
\def\theequation{A-\arabic{equation}} 
The functions $\eta(U)$ and $\cE(y)$ used in the text are
\begin{eqnarray}
\eta(U)&=& e^{\pi i U/12} \prod_{n\geq 1} (1- e^{2 i \pi n U})
\nonumber\\
\nonumber\\
E(y) & = & \frac{\pi y}{ U_2} \ln(4\pi e^{-\gamma} \tau\, U_2^2)
-2\ln \frac{\sinh\pi y^{{1}/{2}}}{\pi y^{{1}/{2}} }
+2\pi^{1/2} U_2 \sum_{k\geq  1}
\frac{\Gamma[k+1/2]}{(k+1)!}\bigg[\frac{-y}{U_2^2}\bigg]^{k+1}\!\!
\zeta[2k+1] 
\nonumber\\
\nonumber\\
& - & 
\ln \prod_{m_1\geq 1}\bigg[
{\bigg\vert 1- e^{-2\pi (y+U_2^2 m_1^2)^{1/2}}\,
 e^{2 i \pi U_1 m_1}\bigg\vert^4}\,\,{\bigg
\vert 1- e^{2 i \pi U m_1}\bigg\vert^{-4}\bigg] }\label{wr2}
\end{eqnarray}
with $E(y\ra 0)\ra 0$.
The function $\cG(q)$ used in the text eq.(\ref{ttt6})
is defined as
\begin{eqnarray}\label{gq1}
\cG(q)& \equiv & 2\ln\pi+2\pi (\lambda \cR)^2
\ln2\pi + 2 \int_{0}^{1} dx\, x\, (1-x)\, \cE\bigg( (R_2 \sin\theta)^2
 (\lambda^2+x(1-x) q^2)\bigg)
\end{eqnarray}
The series of Riemann $\zeta$-functions present in  $E$ (uniformly
 convergent under the conditions of eqs.(\ref{wr1}), (\ref{cgrhtjk})) 
can be integrated termwise.
Removing the IR regulator ($\lambda_0\!\ra\!0$ or
 $\lambda\!\ll\! 1/R_{1,2}^2$) gives
\begin{eqnarray}\label{gq2}
\cG(q)& \equiv & 2\ln\pi+
 2 \int_{0}^{1} dx\, x\, (1-x)\, E\bigg( (R_2 \sin\theta)^2
x(1-x) q^2\bigg)
\end{eqnarray}

\end{document}